\documentclass[aps,pra,onecolumn,superscriptaddress]{revtex4-2}
\pdfoutput=1 

\usepackage[percent]{overpic}
\usepackage{amsmath}
\usepackage{siunitx}
\usepackage{graphicx}
\usepackage{subcaption}
\usepackage{braket}
\usepackage{makecell}
\usepackage{booktabs}
\usepackage{textcomp}
\usepackage{tablefootnote}

\begin{document}

\title{A Fibre-Integrated SPDC Source of Broadband Polarisation-Entangled Photon Pairs at Telecom Wavelengths using Zn-indiffused PPLN Ridge Waveguides}

\author{Peter Iveson}
\email{P.W.Iveson@soton.ac.uk}
\affiliation{Optoelectronics Research Centre, University of Southampton, University Road, Southampton SO17 1BJ, UK}

\author{Paolo L.~Mennea}
\affiliation{Optoelectronics Research Centre, University of Southampton, University Road, Southampton SO17 1BJ, UK}

\author{Goronwy Tawy}
\affiliation{Optoelectronics Research Centre, University of Southampton, University Road, Southampton SO17 1BJ, UK}

\author{Rex H.~S.~Bannerman}
\affiliation{Optoelectronics Research Centre, University of Southampton, University Road, Southampton SO17 1BJ, UK}

\author{Noe Palomar-Davidson}
\affiliation{Optoelectronics Research Centre, University of Southampton, University Road, Southampton SO17 1BJ, UK}

\author{Greg Blanchard-Emmerson}
\affiliation{Covesion Ltd., Unit F3, Adanac Park, Nursling, Southampton SO16 0BT, UK}

\author{Patrick M. Ledingham}
\affiliation{Optoelectronics Research Centre, University of Southampton, University Road, Southampton SO17 1BJ, UK}

\author{Peter G.~R.~Smith}
\affiliation{Optoelectronics Research Centre, University of Southampton, University Road, Southampton SO17 1BJ, UK}

\author{James C.~Gates}
\affiliation{Optoelectronics Research Centre, University of Southampton, University Road, Southampton SO17 1BJ, UK}

\author{Corin B.~E.~Gawith}
\affiliation{Optoelectronics Research Centre, University of Southampton, University Road, Southampton SO17 1BJ, UK}
\affiliation{Covesion Ltd., Unit F3, Adanac Park, Nursling, Southampton SO16 0BT, UK}

\begin{abstract}
We demonstrate a fully fibre-integrated source of broadband polarisation-entangled photon pairs at telecom wavelengths based on a commercially available Zn-indiffused MgO:PPLN ridge waveguide incorporated within a Sagnac interferometer. Operating across the full degenerate SPDC spectrum using pump-suppression filtering only, the source exhibits a measured bandwidth of approximately \SI{67}{\nano\metre}. Strong non-classical correlations are observed, with an average two-photon interference visibility of $96.9\pm0.2$\%, a fidelity to the $\vert{}\Phi^+\rangle$ Bell state of 78.1\%, and a CHSH Bell parameter of $S=2.501\pm0.005$. We further explore the effect of increasing source brightness on Bell-state performance, revealing the limitations of accidental-corrected metrics in high-flux operating regimes.
\end{abstract}

\maketitle

\section{Introduction}
\label{sec:Introduction}

Spontaneous parametric down-conversion (SPDC) presents a highly effective method for producing correlated polarisation-entangled photon pairs, which are foundational to a wide array of applications including quantum communications~\cite{RN324,RN325,RNAlibart2016}, cryptographic protocols~\cite{RN311,RN318}, and quantum metrology~\cite{RN322,RN323,RN321}. To date, a variety of architectures employing SPDC in lithium niobate have been explored to generate these polarisation-entangled states within fibre-integrated platforms~\cite{RN316,RN279,RNVergyris,Yadav2024}.

Single-pass waveguide implementations based on Type-II SPDC naturally generate orthogonally polarised photon pairs~\cite{RN51}. However, the emitted photons are not intrinsically entangled within the nonlinear medium itself~\cite{RN163}. Polarisation entanglement must instead be generated through probabilistic post-selection, an approach that typically incurs a minimum 50\% reduction in flux~\cite{RN159}. While deterministic post-selection schemes exist, they generally produce narrow-band, non-degenerate spectra and necessitate complex engineering interventions, such as interlaced dual poling periods~\cite{RN33,RN157,RN314}. Furthermore, these approaches require strict compensation for the temporal walk-off between the orthogonally polarised modes, as well as precise phase control to maintain indistinguishability between interferometric pathways. While temporal walk-off control has been demonstrated in fibred setups~\cite{RN316}, it implicitly relies on aggressive spectral filtering, limiting the useful output bandwidth.

Additionally, Type-II processes in lithium niobate typically access the off-diagonal nonlinear tensor elements, which are substantially weaker than the dominant $d_{33}$ nonlinear coefficient accessible to Type-0 processes. This results in an order-of-magnitude reduction in nonlinear interaction efficiency~\cite{RN326,RN18}. Furthermore, Type-II SPDC is often constrained in bandwidth by the group-velocity mismatch between the orthogonal modes~\cite{RN327}. In contrast, Type-0 SPDC benefits from reduced mismatch in co-polarised interactions, enabling the broad spectral emission highly sought after for wavelength-multiplexed quantum communication architectures~\cite{Yadav2022}. However, because the signal and idler photons in a Type-0 process share the same polarisation, alternative interferometric techniques are required to generate entanglement. 

Early demonstrations of Type-0 polarisation entanglement employed dual-waveguide or Mach-Zehnder interferometric schemes~\cite{RN279}, which suffered from stringent phase-stabilisation requirements. More recently, Sagnac interferometer configurations have emerged as a robust alternative. Owing to their common-path geometry, Sagnac loops exhibit intrinsic phase stability and do not require active feedback control~\cite{RN328}. This topology has been successfully demonstrated in both free-space and guided-wave implementations~\cite{RN329,RNVergyris,Yadav2024}.

\begin{figure}[t]
  \centering
  \includegraphics[width=6cm]{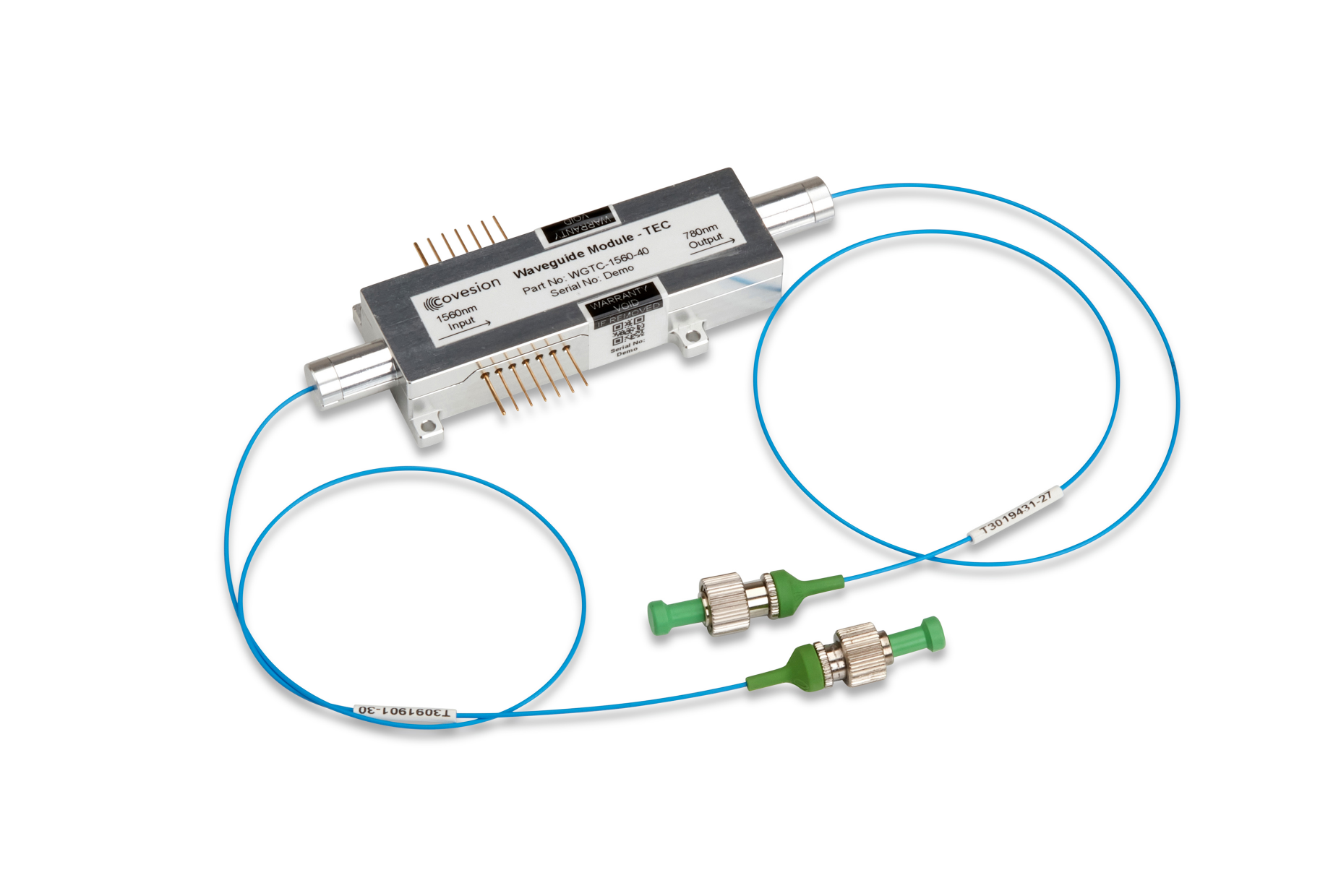}
  \caption{Covesion Ltd. fibre-pigtailed waveguide module (Figure courtesy of Covesion Ltd.).}
  \label{fig:waveguidemodule}
\end{figure}

In this work, we demonstrate a broadband fibre-integrated source of polarisation-entangled photon pairs based on a commercially available pigtailed Zn-indiffused MgO:PPLN ridge waveguide operated within a Type-0 Sagnac interferometer. The system is realised entirely using off-the-shelf commercial fibre components and a pigtailed Zn-indiffused MgO:PPLN ridge waveguide module (Covesion Ltd., Fig.~\ref{fig:waveguidemodule}). Unlike previous fibre implementations relying on deterministic spectral separation near degeneracy, the source operates across the full degenerate SPDC bandwidth ($\sim\SI{67}{\nano\metre}$) using only pump-suppression filtering. We characterise the source through spectral measurements, two-photon interference, CHSH Bell tests and quantum state tomography. Finally, we investigate the dependence of Bell-state performance on source brightness, revealing the increasing influence of accidental and multi-pair contributions at high coincidence rates.

\section{SHG and spectral characterisation}
\label{sec:characterisation}
Our SPDC device is a Zn‑indiffused, MgO‑doped PPLN waveguide (Covesion) with ridge width \SI{11.7}{\micro\metre}, height \SI{30}{\micro\metre}, and length \SI{40}{\milli\metre}, pigtailed with \SI{1550}{\nano\metre} PM fibre at both ends. The waveguide fabrication process and structure are similar to those described in \cite{RNCarpenter,RN379}. The on‑module oven provides temperature control. The structure is designed for Type‑0 \SI{1560}{\nano\metre} $\rightarrow$ \SI{780}{\nano\metre} QPM ($e \rightarrow e + e$) and here is used as such for SHG, and in reverse (SPDC: \SI{780}{\nano\metre} $\rightarrow$ \SI{1560}{\nano\metre}); a nominal poling period of \SI{18.6}{\micro\metre} is required~\cite{RNCarpenter}. 

\subsection{Second‑harmonic generation (SHG)}
\label{sec:SHG}
\begin{figure}[t]
  \centering
  \begin{subfigure}[t]{0.4\linewidth}
    \centering
    \caption*{(a)}
    \includegraphics[width=\linewidth]{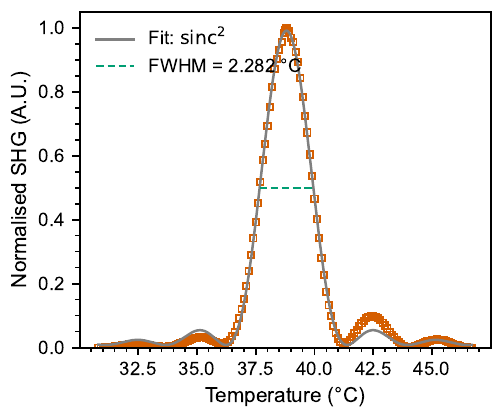}
    \label{fig:shg_2panel_temp}
  \end{subfigure}\hfill
  \begin{subfigure}[t]{0.39\linewidth}
    \centering
    \caption*{(b)}
    \includegraphics[width=\linewidth]{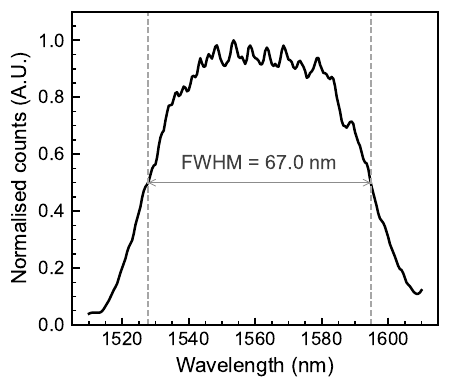}
    \label{fig:spdc_spectrum}
  \end{subfigure}
  \caption{SHG characterisation and SPDC spectral response. (a) SHG temperature acceptance. Fit yields $T_0=\SI{38.81}{\celsius}$, $\mathrm{FWHM}_T=\SI{2.282}{\celsius}$. Markers, data; orange curve, $\mathrm{sinc}^2$ fit; green dashed line, FWHM indication. (b) SPDC degenerate spectrum measured at $T=\SI{38.37}{\celsius}$.}
  \label{fig:shg2_panel}
\end{figure}

We use SHG to locate phase‑matching, exploiting the higher conversion efficiency and lower noise than SPDC. Experimentally, using a narrow‑linewidth C‑band laser (Santec WSL‑110), we measure $T_0=\SI{38.81}{\celsius}$ and $\mathrm{FWHM}_T=\SI{2.282}{\celsius}$ at $\lambda_p=\SI{1560}{\nano\metre}$ (Fig.~\ref{fig:shg2_panel}(a)).  A typical  $\mathrm{sinc}^2$ shape is apparent indicating excellent poling homogeneity. The slight asymmetry in our experimental sidelobes may be due to a small variation in ridge width \cite{RNGrayvariations}. 

\begin{figure*}[t]
  \centering
  \begin{subfigure}[t]{0.90\textwidth}
    \centering
    \caption*{(a)}
    \includegraphics[width=\linewidth]{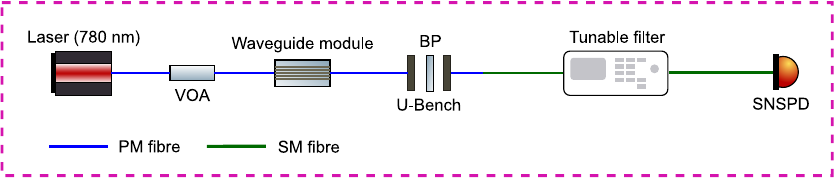}
    \label{fig:spdc_generation}
  \end{subfigure}

  \begin{subfigure}[t]{0.9\textwidth}
    \centering
    \caption*{(b)}
    \includegraphics[width=\linewidth]{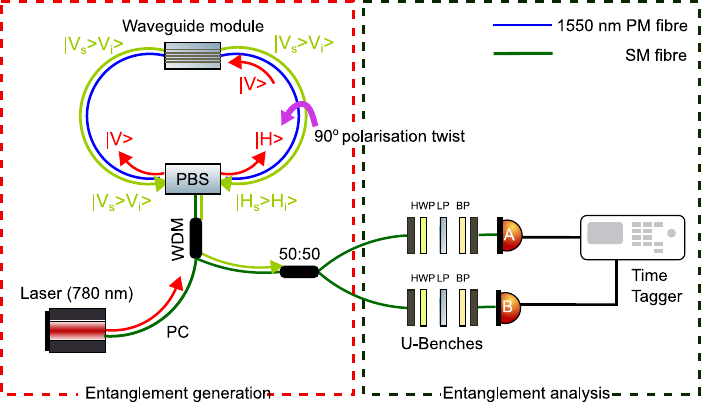}
    \label{fig:car_setup}
  \end{subfigure}
  \

  \caption{Overview of the SPDC configurations. 
  (a) Spectral characterisation with a tunable filter. 
  (b) Experimental setup for polarisation entanglement in a Sagnac configuration using commercial components throughout.  PM: polarisation-maintaining. SM: single-mode, PBS: fibre polarising beam splitter, VOA: variable optical attenuator, PC: Fibred polarisation controllers, WDM: Wavelength-division-multiplexer, BP: 1100 nm Longpass filter (broadband entanglement) / $1560 \pm 6$ nm Bandpass filter (narrowband entanglement). A: Alice, B:Bob SNSPD channels.
 }
  \label{fig:main_2x2}
\end{figure*}

\subsection{SPDC bandwidth measurement}
\label{sec:SPDC_bandwidth_measurement}

To generate the SPDC output, the pigtailed Covesion PPLN ridge module was pumped at \SI{780}{\nano\metre}. The setup is shown in Fig.~\ref{fig:main_2x2}(a). The pump was a continuous-wave PM-fibred single-mode laser (Thorlabs DBR780PN, nominal linewidth \SI{1}{\mega\hertz}), with a variable optical attenuator (Thorlabs VOA780PM-APC) to set the input power. 

The degenerate SPDC output near \SI{1560}{\nano\metre} was delivered by fibre to a 16-channel superconducting nanowire single-photon detector (SNSPD) system (ID Quantique ID281). A single SNSPD channel was utilised for these measurements, with an efficiency of 84\%, timing jitter of \SI{31}{\pico\second}, dead time of \SI{34}{\nano\second}, and a dark-count rate of 22 cps.  Time tagging used a Swabian TimeTagger with Python-based analysis.

Residual pump light was rejected using a free-space silicon filter (Thorlabs FBHQ1550), providing approximately \SI{140}{\decibel} suppression, including measured optical-path attenuation. To verify pump suppression, we detuned from phase matching and observed that detector count rates fell to background. Background count rates, measured with the pump extinguished, were comparable to the detector dark-count rates under operating conditions. 

For bandwidth characterisation, the SPDC output was filtered using a manually tunable filter (Alnair BVF-100). The phase-matching temperature was adjusted slightly from $T_0=\SI{38.83}{\celsius}$ to $T_0=\SI{38.37}{\celsius}$ to achieve degeneracy at $\sim\SI{1560}{\nano\metre}$. The centre wavelength was scanned across the C- and L-bands ($\approx\SIrange{1510}{1610}{\nano\metre}$) at the phase-matching temperature, with a sub-nanometre passband. Normalised counts are shown in Fig.~\ref{fig:shg2_panel}(b). The measured broadband spectrum has a bandwidth of $\sim\SI{67}{\nano\metre}$ (FWHM) and shows potential for wavelength-multiplexed quantum networking.

\section{Polarisation entanglement}
\label{sec:polarisation_entanglement}

\subsection{Experimental setup}
\label{sec:Experimental_setup}

Following~\cite{RNVergyris,Yadav2024}, the waveguide module was incorporated into a fibre-based Sagnac interferometer configuration (Fig.~\ref{fig:main_2x2}(b)). As above, the system was pumped by a PM-fibred continuous-wave \SI{780}{\nano\metre} laser (Thorlabs DBR780PN), with a variable optical attenuator (Thorlabs VOA780PM-APC) used to control the input power.

In the Sagnac configuration, the \SI{780}{\nano\metre} pump field is split using a fibre polarising beam splitter into counter-propagating paths, one of which undergoes a $90^{\circ}$ polarisation rotation. The waveguide is subsequently pumped from both ends by a coherent superposition of orthogonally polarised fields, resulting in the generation of the entangled state
\begin{equation}
\ket{\Phi_{\mathrm{out}}} = \frac{1}{\sqrt{2}}\left(\ket{HH}_{\mathrm{Path1}} + \ket{VV}_{\mathrm{Path2}}\right).
\end{equation}
Recombination of the generated photon amplitudes at the polarising beam splitter erases path information, yielding the polarisation Bell state
\begin{equation}
\ket{\Phi^+} = \frac{1}{\sqrt{2}}\left(\ket{HH} + \ket{VV}\right).
\end{equation}

The output photons are subsequently separated using a  $1\times2$ fibre coupler (Thorlabs TW1550R5F1) into signal and idler channels. The use of a broadband 50:50 coupler provides wavelength-independent photon separation, allowing potential operation at different degenerate wavelengths without modification of the separation hardware. In the analysis stage, the nominated signal (Alice) and idler (Bob) photons are projected onto polarisation bases using linear polarisers (LPs), with measurement settings controlled by half-wave plates (HWPs). Alice’s analyser is fixed while Bob’s HWP is scanned to obtain interference fringes. Coincidence and singles rates are recorded using two channels of the SNSPD system. The utilised detector channels exhibit efficiencies of 84\% and 90\%, timing jitters of \SI{29}{\pico\second} and \SI{27}{\pico\second}, dead times of \SI{38}{\nano\second} and \SI{30}{\nano\second}, and dark-count rates of 20 cps and 30 cps, respectively. Time tagging was performed using a Swabian TimeTagger, with data processing implemented in Python. Measurements were performed in a broadband configuration using longpass filters (Thorlabs FELH1100) providing approximately \SI{120}{\decibel} of pump suppression, including measured filter and optical path attenuation. Additional spectral filtering was introduced, where required, by substituting the longpass filters with \SI{1560 \pm 6}{\nano\metre} bandpass filters (Thorlabs FBH1560-12). The use of free-space dielectric filters thus provides an easily reconfigurable approach to assess system performance under partial spectral selection.

\subsection{Two-photon interference and visibilities}
\label{sec:two_photon_interference}

\begin{figure}[htbp]
  \centering
  
  \begin{subfigure}{0.96\linewidth}
    \caption{} 
    \includegraphics[width=\linewidth]{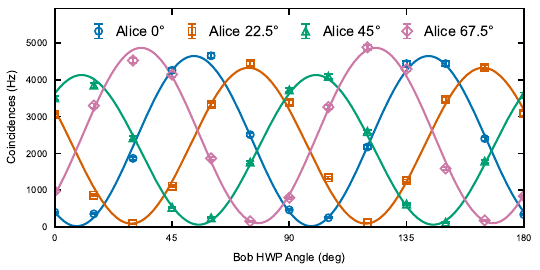}
    \label{fig:visibilities_all}
  \end{subfigure}
  
  
  \begin{subfigure}{0.5\linewidth}
    \caption{} 
    \includegraphics[width=\linewidth]{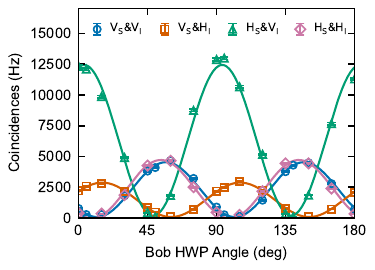}
    \label{fig:visibilities_alice_0}
  \end{subfigure}
  \hfill 
  \begin{subfigure}{0.5\linewidth}
    \caption{} 
    \includegraphics[width=\linewidth]{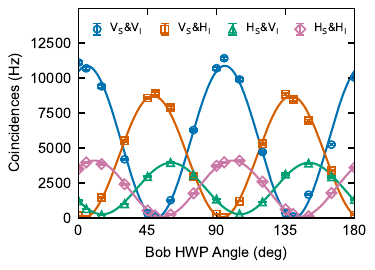}
    \label{fig:visibilities_alice_45}
  \end{subfigure}
  
  \caption{Coincidence counts: (a) the Alice photon is projected in a linear polarisation basis and the Bob photon HWP angle is scanned. LP is set to H ($90^\circ$) for all measurements. (b) Visibility curves for Alice set to $0^\circ$. (c) Visibility curves for Alice set to $45^\circ$. Markers show the measured coincidence counts and the solid line gives the sinusoidal fit. Peaks are phase-shifted relative to the canonical $|\Phi^+\rangle$ expectation. Error bars assume Poissonian statistics. Legends in (b) and (c) show LP settings. LP: Linear polariser.}
  \label{fig:visibilities}
\end{figure}

Strong two-photon interference fringes were observed across multiple mutually unbiased bases during fully broadband operation (employing only longpass pump suppression) (Fig.~\ref{fig:visibilities}(a)). The interference fringe visibility is defined as 
\begin{equation}
V = \frac{C_{\max} - C_{\min}}{C_{\max} + C_{\min}},
\end{equation}
where \(C_{\max}\) and \(C_{\min}\) denote the maximum and minimum coincidence counts extracted from a Poisson-weighted non-linear least-squares fit to the experimental data. To ensure physically meaningful results in the presence of Poissonian counting fluctuations near the coincidence minima, the fit was constrained to the range \(0 \leq V \leq 1\).Visibilities of \qty[separate-uncertainty=true]{99.0(4)}{\percent},  \qty[separate-uncertainty=true]{95.8(3)}{\percent}, \qty[separate-uncertainty=true]{97.0(4)}{\percent}, and \qty[separate-uncertainty=true]{95.8(4)}{\percent} were measured in the H ($0^\circ$), V ($45^\circ$), D ($22.5^\circ$), and A ($67.5^\circ$) projections of the signal photon respectively, yielding an average visibility of  \qty[separate-uncertainty=true]{96.9(2)}{\percent}. Singles rates remained relatively stable throughout the acquisition, confirming that the high fringe contrast arises from non-local correlations rather than source intensity fluctuations. Reported coincidence rates correspond to measured detection rates. Owing to the probabilistic nature of the 50:50 splitting stage, the generated pair rate is expected to be approximately twice the measured coincidence rate \cite{RN257}.

To further verify the entangled nature of the state, interference fringes were mapped across four orthogonal polariser combinations ($\mathrm{V_S}$\&$\mathrm{V_I}$, $\mathrm{V_S}$\&$\mathrm{H_I}$, $\mathrm{H_S}$\&$\mathrm{V_I}$, and $\mathrm{H_S}$\&$\mathrm{H_I}$). Fig.~\ref{fig:visibilities}(b and c) illustrate the recorded coincidences for signal analyser settings of $0^\circ$ and $45^\circ$, respectively. As expected, the co-polarised modes ($\mathrm{V_S}$\&$\mathrm{V_I}$, $\mathrm{H_S}$\&$\mathrm{H_I}$) and cross-polarised modes ($\mathrm{H_S}$\&$\mathrm{V_I}$, $\mathrm{V_S}$\&$\mathrm{H_I}$) exhibit corresponding in-phase and anti-phase sinusoidal behaviours. A slight angular offset ($\lesssim 12^\circ$) observed between the fitted cross-polarised fringes is attributed primarily to the mechanical precision of the analyser components and finite sampling resolution. The fitted maxima of the orthogonal analyser settings are separated by $\sim46^\circ$, showing excellent agreement with the theoretical $45^\circ$ expectation. Crucially, while rotating the signal analysis basis shifts the absolute phase of the coincidences, the relative phase difference between the individual complementary fringes is preserved. This signature serves as a classical witness to the generation of a maximally entangled Bell state~\cite{RNVergyris}.

Fringe visibilities for the four polariser settings with the signal photon projected at $0^\circ$ (Fig.~\ref{fig:visibilities_alice_0}) were evaluated as \qty[separate-uncertainty=true]{95.6(3)}{\percent}, \qty[separate-uncertainty=true]{93.8(4)}{\percent}, \qty[separate-uncertainty=true]{98.5(1)}{\percent}, and \qty[separate-uncertainty=true]{99.0(4)}{\percent}, respectively. For a signal projection of $45^\circ$ (Fig.~\ref{fig:visibilities_alice_45}), the corresponding visibilities were \qty[separate-uncertainty=true]{98.5(1)}{\percent}, \qty[separate-uncertainty=true]{98.6(1)}{\percent}, \qty[separate-uncertainty=true]{88.0(5)}{\percent}, and \qty[separate-uncertainty=true]{96.7(3)}{\percent}. We note a variation in the absolute coincidence count maxima between different measurement bases. This amplitude asymmetry is ascribed to a slight imbalance in the input pump polarisation, leading to unequal coherent superposition amplitudes within the Sagnac loop, together with the inherent polarisation-dependent detection efficiency of the SNSPDs. These amplitude disparities do not degrade the underlying interference visibility and can be mitigated in future iterations through strict pre-compensation of the pump polarisation and the integration of fibre polarisation controllers immediately preceding the detector inputs.

\subsection{CHSH inequality violation}

The violation of the Clauser--Horne--Shimony--Holt (CHSH) inequality provides a definitive metric for testing non-local hidden variables and determining the degree of source entanglement~\cite{RN317}. To quantify this, we evaluate the CHSH Bell parameter $S$ following~\cite{RN320,WallsMilburn}:
\begin{equation}
S = \left| E(a,b) - E(a,b') + E(a',b) + E(a',b') \right|,
\end{equation}
where $E(a,b)$ represents the correlation function measured for the analyser half-wave plate (HWP) settings $a$ and $b$ on the signal (Alice) and idler (Bob) photons, respectively. 

The correlation function is extracted directly from the coincidence measurements according to
\begin{equation}
E(a,b) = \frac{C_{++} + C_{--} - C_{+-} - C_{-+}}{C_{++} + C_{--} + C_{+-} + C_{-+}},
\end{equation}
where $C_{ij}$ denotes the recorded coincidence count for the projection outcomes $i$ and $j$ in a given measurement basis.

Quantum mechanics dictates a maximum Bell parameter of $S = 2\sqrt{2} \approx 2.82$ for a maximally entangled $\ket{\Phi^+}$ state at the canonical measurement angles $a=0^\circ$, $a'= 22.5^\circ$, $b=11.25^\circ$, and $b'= 33.75^\circ$. Any classical local hidden variable theory is strictly bounded by $S \leq 2$.

Because the single-mode fibre pigtails impart arbitrary local unitary rotations on the distributed photons, the required hardware measurement angles must be deterministically extracted from the phase shifts of the measured two-photon interference fringes. Operating in the fully broadband configuration (pump suppression only) at a raw coincidence rate of $\sim5\times10^{3}$ Hz the system yielded $S = 2.501 \pm 0.005$. This represents a violation of the classical CHSH inequality by $\sim100$ standard deviations, confirming the presence of strong non-classical correlations across the full degenerate SPDC bandwidth.

Substituting the broadband longpass filters with a spectral bandpass filter ($1560 \pm 6$~nm) increased the measured Bell parameter to $S = 2.66 \pm 0.03$ at a raw coincidence rate of $\sim1.5\times10^{2}$ Hz. This enhancement in $S$ under spectral filtering is attributed to the suppression of wavelength-dependent phase dispersion across the broad emission spectrum. Crucially, the observation of a definitive CHSH inequality violation without narrowband filtering demonstrates that the Sagnac source maintains high-quality entanglement even when operating in a fully broadband regime.

\subsection{Brightness-dependent Bell state performance}
\label{sec:Bell_state_performance}

To evaluate the robustness of the entanglement against increasing source brightness, the CAR and the Bell parameter were recorded over a broad range of detected coincidence rates. The maximum recorded coincidence rate of roughly 121 kHz was dictated by the maximum available pump power (\(<\SI{0.25}{\milli\watt}\)) delivered into the Sagnac loop, and corresponds to an estimated pair generation rate of $\sim2.7$ MHz at the output of the pigtailed waveguide module after accounting for downstream optical losses, detector efficiencies and the probabilistic 50:50 splitting stage.

We quantify pair correlations using the coincidences-to-accidentals ratio (CAR)~\cite{Yadav2022}, \begin{equation} \mathrm{CAR} = \frac{C_T}{C_A}, \end{equation} where \(C_T\) is the zero-delay coincidence count and \(C_A\) is the accidental estimate, obtained as the mean coincidence rate well beyond the coherence time multiplied by the coincidence-window width.

\begin{figure}[t]
  \centering
    \begin{subfigure}[t]{0.48\linewidth} 
    \centering
    \caption*{(a)}
    \includegraphics[width=\linewidth]{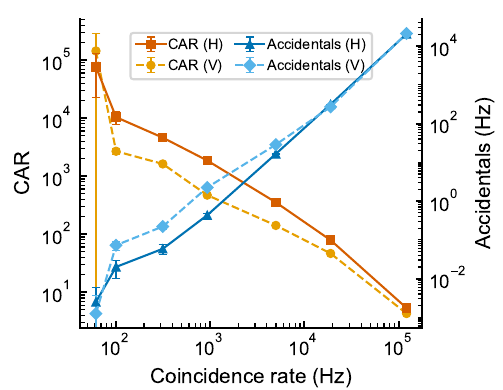}
    \label{fig:cars_and_accidentals}
  \end{subfigure}
  \hfill
    \begin{subfigure}[t]{0.48\linewidth} 
    \centering
    \caption*{(b)}
    \includegraphics[width=\linewidth]{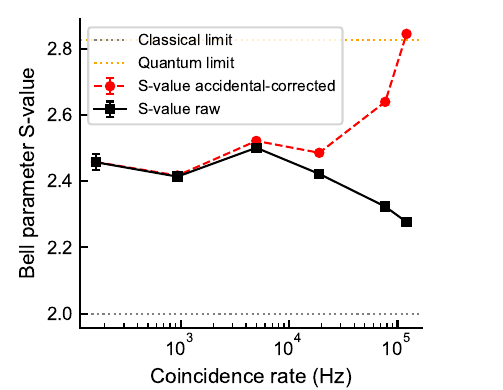}
    \label{fig:s_values}
  \end{subfigure}
\caption{Dependence of entanglement metrics on raw coincidence rate for broadband operation.
(a) Measured CAR values and accidental coincidence rates in the H and V polarisation bases as a function of raw coincidence rate.
(b) Raw and accidental-corrected CHSH Bell parameter as a function of raw coincidence rate. Error bars correspond to one standard deviation obtained from propagated counting statistics.}
  \label{fig:s_values_and_car}
\end{figure}

Fig.~\ref{fig:s_values_and_car}(a) shows the CAR and accidental events as a function of the coincidence rate. As the source brightness increases, the CAR undergoes an expected reduction, falling from $>7.6 \times 10^4$ to $\approx 5$ at the highest investigated operating point. This behaviour is intrinsically linked to the increased statistical probability of multipair emissions and uncorrelated background events falling within the temporal coincidence window  \cite{RN338}.

The corresponding impact of source brightness on the CHSH Bell parameter is shown in Fig.~\ref{fig:s_values_and_car}(b). To isolate the influence of the rising background, accidental-corrected Bell parameters ($S_{\text{corr}}$) were calculated alongside the raw data. At coincidence rates below the 5 kHz regime, the raw and accidental-corrected $S$-values are highly aligned, exhibiting negligible variance and confirming that the intrinsic state purity remains high. However, as the coincidence rate scales toward the 121 kHz limit, a clear divergence emerges. The raw Bell parameter steadily degrades to $S_{\text{raw}} = 2.276 \pm 0.001$ suggesting state dilution. Conversely, applying a standard accidentals subtraction at this extreme brightness inflates the corrected value to $S_{\text{corr}} = 2.845 \pm 0.001$, exceeding Tsirelson's bound for bipartite entanglement ($2\sqrt{2} \approx 2.828$) \cite{RN339}.  Similar behaviour has been reported in previous SPDC Bell-test experiments in which accidental-corrected Bell parameters exceeded the quantum limit \cite{RN341}. This indicates that while accidental subtraction can provide a useful estimate of the underlying source performance at low pair-production rates, the resulting Bell parameters should be interpreted with caution in high-brightness regimes where accidental and multi-pair contributions become significant.

\section{Quantum state tomography}

\begin{figure}[t]
  \centering
  
  \begin{subfigure}[t]{0.4\linewidth}
    \centering
    \caption*{(a)}
    \includegraphics[width=\linewidth]{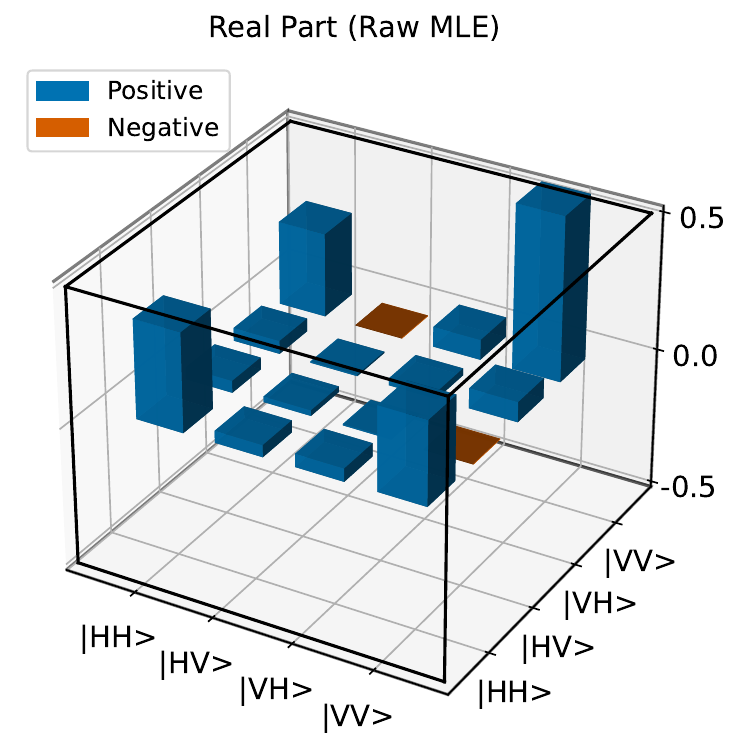}
    \label{fig:shg_2panel:a}
  \end{subfigure}\hspace{0.8cm} 
  \begin{subfigure}[t]{0.4\linewidth}
    \centering
    \caption*{(b)}
    \includegraphics[width=\linewidth]{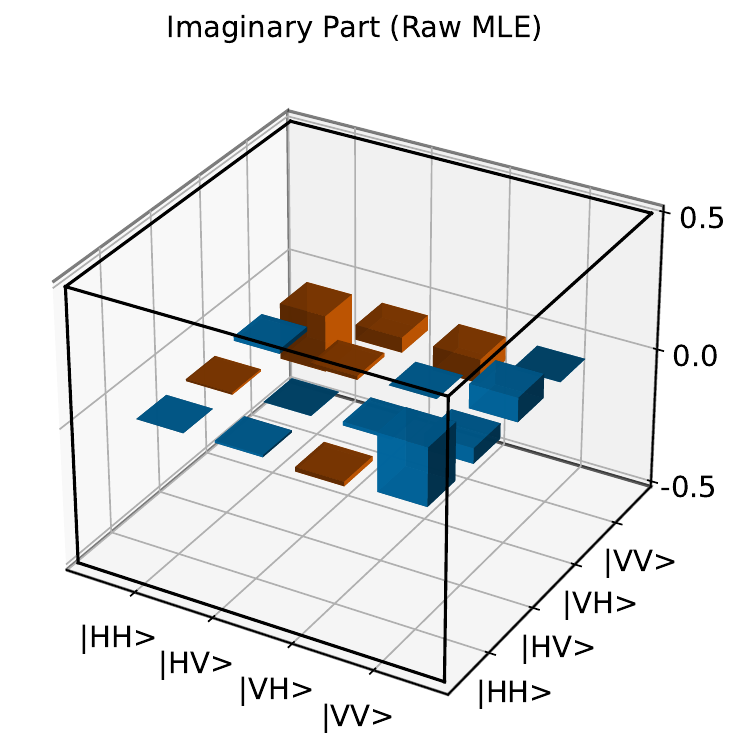}
    \label{fig:shg_2panel:b}
  \end{subfigure}
  
  
  \begin{subfigure}[t]{0.4\linewidth}
    \centering
    \caption*{(c)}
    \includegraphics[width=\linewidth]{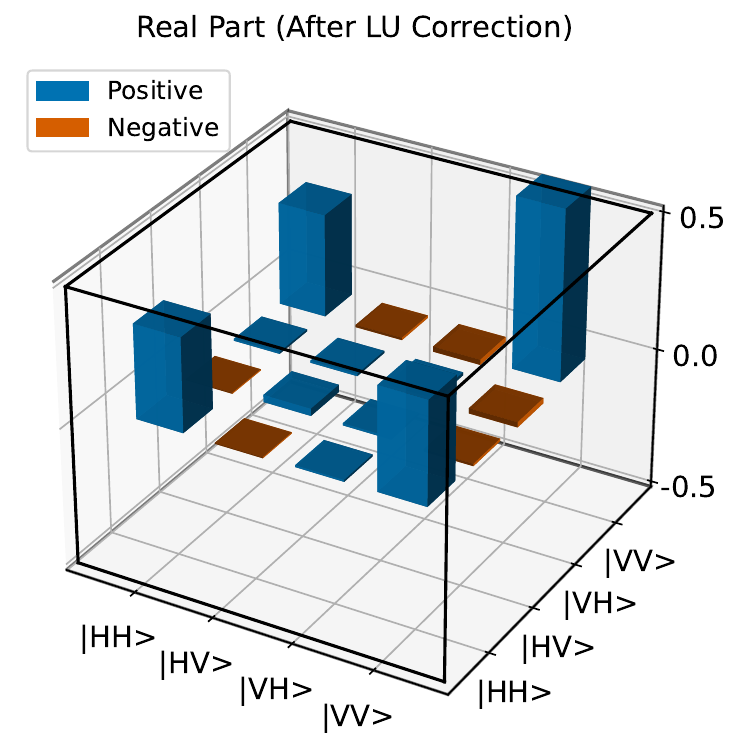}
    \label{fig:shg_2panel:c} 
  \end{subfigure}\hspace{1cm} 
  \begin{subfigure}[t]{0.4\linewidth}
    \centering
    \caption*{(d)}
    \includegraphics[width=\linewidth]{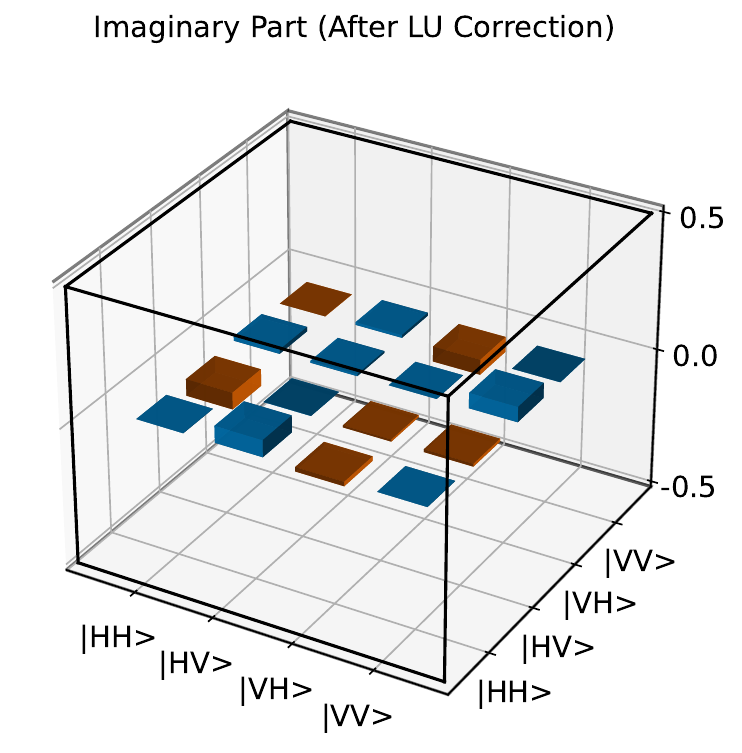}
    \label{fig:shg_2panel:d} 
  \end{subfigure}
  
  \caption{Reconstructed density matrices. (a) Real and (b) imaginary components of the raw density matrix yielding a fidelity of $78.1\%$ to the ideal $\vert{}\Phi^+\rangle$ Bell state. (c) Real and (d) imaginary components of the optimised density matrix, improving the fidelity to $85.6\%$.}
  \label{fig:qst_4panel}
\end{figure}

To fully characterise the generated state and quantify its overlap with an ideal Bell state, we reconstructed the two-qubit density matrix via quantum state tomography (QST)~\cite{RN301}. Coincidence measurements were performed using the broadband configuration detailed in Fig.~\ref{fig:main_2x2}(b), supplemented by a quarter-wave plate (QWP) preceding the HWP in both analyser arms to access the right- and left-circular (R and L) polarisation bases. A complete set of 16 projection measurements was acquired across the mutually unbiased H-V, D-A, and R-L bases.

Fig.~\ref{fig:qst_4panel}(a and b) present the real and imaginary components of the reconstructed raw density matrix. From this we calculate a raw state fidelity of 78.1\% (extracted directly from raw coincidence counts without background accidental subtraction via maximum likelihood estimation) with respect to the ideal maximally entangled Bell state $\ket{\Phi^+} = \frac{1}{\sqrt{2}}\left(\ket{HH} + \ket{VV}\right)$.  The measured fidelity exceeds the separability threshold of $\approx0.71$ for the Bell state $\ket{\Phi^+}$, thereby confirming the presence of bipartite entanglement~\cite{RN360}.

Because the single-mode fibre pigtails introduce arbitrary local unitary rotations on the distributed photons, the reconstructed density matrix need not be expressed in the natural basis of the generated Bell state. Following \cite{RN331}, local unitary optimisation was therefore applied to identify the single-qubit transformations required to maximise the overlap of the reconstructed state with the ideal $\ket{\Phi^{+}}$ Bell state. This procedure increased the fidelity from 78.1\% to 85.6\%, consistent with the high interference visibilities and CHSH Bell parameters reported. The improvement suggests that residual local basis rotations, rather than a lack of entanglement, are a significant contributor to the reduced raw fidelity. Future work will investigate the implementation of the corresponding compensating waveplate settings directly within the analyser hardware.

To provide a state-agnostic metric of intrinsic entanglement, we additionally evaluate the concurrence $C(\rho)$, defined as~\cite{RN330}:
\begin{equation}
C(\rho) = \max \left( 0,\, \lambda_1 - \lambda_2 - \lambda_3 - \lambda_4 \right),
\end{equation}
where $\lambda_i$ are the square roots of the eigenvalues, in decreasing order, of the matrix
\begin{equation}
R = \rho\, (\sigma_y \otimes \sigma_y)\, \rho^*\, (\sigma_y \otimes \sigma_y).
\end{equation}
Here, $\rho^*$ denotes the complex conjugate of the density matrix $\rho$ in the computational basis, and $\sigma_y$ is the Pauli-$y$ matrix. We observe a concurrence of 0.76. Together with the measured raw fidelity, high interference visibilities and CHSH Bell parameter significantly exceeding the classical limit, this result provides strong evidence for the generation of a high-quality broadband polarisation-entangled state. The overall source performance is summarised and compared with representative published fibre-based Type-0 PPLN Sagnac architectures in Table~\ref{tab:spdc-sources}.

\begin{table}[h!] \centering \small \caption{Comparison of fibred Type-0 Sagnac loops in the telecom band in PPLN waveguides.} \label{tab:spdc-sources} \setlength{\tabcolsep}{4pt} \begin{tabular}{lccccccc} \toprule \textbf{Source} & \makecell[t]{\textbf{Filter}\\\textbf{BW}\\(nm)} & \makecell[t]{\textbf{Brightness}\\(pairs/s/mW/GHz)} & \makecell[t]{\textbf{Average}\\\textbf{visibility} \\(\%)} & \makecell[t]{\textbf{Fidelity}} & \makecell[t]{\textbf{S-value}} & \makecell[t]{\textbf{Source}\\\textbf{BW}\\(nm)} & \makecell[t]{\textbf{Detectors used}\\\textbf{(Q.E.) \& CDW}} \\ \midrule This work & 67 & $1.06 \times 10^{6}$\textsuperscript{a} & 96.9 & 78.1 & $2.501 \pm 0.005$ & 67 & \makecell[t]{SNSPD (86\%\textsuperscript{f})\\600 ps} \\ \cite{Yadav2024} & 52\textsuperscript{b} & $1.6 \times 10^{6}$\textsuperscript{c} & 92 & 89.4 & $2.49 \pm 0.017$ & 52 & \makecell[t]{APD (10\%)\\512 ps} \\ & 0.8 & -- & 93.5 & 96.5 & $2.55 \pm 0.024$ & -- & \makecell[t]{APD (20\%)\\512 ps} \\ \cite{RNVergyris} & 1 & $1.96 \times 10^{6}$ & 85.2 & -- & $2.50 \pm 0.02$ & 40 & \makecell[t]{APD (20\%)\\81 ps} \\ \cite{RN297} & SC & -- & 99.0 & 98.1 & $2.76 \pm 0.03$ & 60 & \makecell[t]{SNSPD (80\%)\\500 ps} \\ \cite{Li:15} & SC & $1.0 \times 10^{6}$\textsuperscript{d} & 88.6 & -- & $2.50 \pm 0.12$ & 3.2 & \makecell[t]{APD (20\%)\\0.4 ns} \\ \cite{RN336} & -- & -- & 93.5\textsuperscript{e} & -- & -- & -- & \makecell[t]{APD (16.3\%)\\--} \\ \bottomrule \end{tabular} \vspace{4pt} 
\begin{minipage}{0.95\linewidth}
\raggedright \footnotesize{ Missing values are indicated by ``--''. BW: bandwidth, SC: single channel, Q.E.: quantum efficiency, CDW: coincidence detection window. APD: avalanche photodiode.\\ \textsuperscript{a} As reported in \cite{RN379}.\\ \textsuperscript{b} Assumed from reported configuration and source bandwidth.\\ \textsuperscript{c} Converted from the reported $2 \times 10^8$ pairs/s/mW/nm using $1\,\mathrm{nm}\approx125\,\mathrm{GHz}$ at $1550\,\mathrm{nm}$.\\ \textsuperscript{d} Reported as pairs/s/100 GHz; not normalised to pump power.\\ \textsuperscript{e} Value after accidentals correction.\\ \textsuperscript{f} Average SNSPD detection efficiency. } \end{minipage} \end{table}

\section{Conclusion} 
We have demonstrated a fully fibre-integrated Type-0 Sagnac source of broadband polarisation-entangled photon pairs based on a commercially available Zn-indiffused MgO:PPLN ridge waveguide. Operating across the full \SI{67}{\nano\metre} degenerate SPDC spectrum using pump-suppression filtering alone, the source exhibited a high average visibility of $96.9\pm0.2$\%, a raw Bell parameter of $S = 2.501 \pm 0.005$, and a concurrence of 0.76. These results demonstrate that high-quality entanglement can be maintained without resorting to aggressive spectral filtering or deterministic spectral partitioning, whilst preserving the broad bandwidth desirable for wavelength-multiplexed quantum communication systems.

The source remained robust over a wide operating range, maintaining Bell inequality violations well above the classical limit while the coincidence-to-accidentals ratio decreased by more than four orders of magnitude. At the highest investigated brightness, the divergence between raw and accidental-corrected Bell parameters highlighted the limitations of simple accidental-subtraction procedures in high-flux SPDC regimes and emphasised the increasing influence of accidental and multi-pair contributions on inferred entanglement metrics.

Overall, the demonstrated architecture combines broadband operation, strong non-classical correlations and implementation using commercially available fibre-coupled components. These characteristics make it an attractive platform for future wavelength-multiplexed quantum networking and scalable fibre-based quantum communication applications.

\begin{acknowledgments}
\textbf{Funding.} UK Quantum Technology Hub in Sensing, Imaging and Timing [QuSIT], (EP/Z533166/1); UK Quantum Technology Hub in Quantum Imaging [QuantIC], (EP/T00097X/1); Engineering and Physical Sciences Research Council, (EP/W524621/1).

\textbf{Disclosures.} The authors declare no conflicts of interest.

\textbf{Data availability.} All data supporting this study are available from the University of Southampton repository and will be made publicly available upon publication.
\end{acknowledgments}
\bibliographystyle{apsrev4-2}
\bibliography{references}
\end{document}